\documentclass[lettersize,journal]{IEEEtran}
\usepackage{amsmath,amsfonts}
\usepackage{algorithmic}
\usepackage{algorithm}
\usepackage{array}
\usepackage{textcomp}
\usepackage{stfloats}
\usepackage{url}
\usepackage{CJKutf8}
\usepackage{verbatim}
\usepackage{graphicx}
\usepackage{cite}

\usepackage{subfigure}
\usepackage{dutchcal} 
\usepackage{amssymb}

\usepackage{enumitem}

\usepackage{booktabs}

\usepackage{multirow}

\usepackage{makecell}

\usepackage{newfloat}
\usepackage{listings}

\usepackage{caption}
\usepackage{subcaption}
\usepackage{arydshln}
\usepackage{booktabs}
\usepackage{graphicx}

\begin{document}

\title{Towards Alleviating Text-to-Image Retrieval Hallucination for CLIP in Zero-shot Learning}

\author{
Hanyao Wang, Yibing Zhan,~\IEEEmembership{Member,~IEEE}, Liu Liu, Liang Ding, Yan Yang, Jun Yu,~\IEEEmembership{Senior Member,~IEEE}

\thanks{
Hanyao Wang, Yan Yang and Jun Yu are with the School of Computer Science, Hangzhou Dianzi University, Hangzhou 310018, China. (e-mail: wanghanyao@hdu.edu.cn, yangyan@hdu.edu.cn, yujun@hdu.edu.cn)\par
Yibing Zhan is with the JD Explore Academy, China. (e-mail: zhanyibing@jd.com)\par
Liu Liu is with the Institute of Artificial Intelligence, Beihang University, Beijing 100191, China (e-mail: liuliubh@gmail.com)\par
Liang Ding is with the University of Sydney, Australia (e-mail: liangding.liam@gmail.com)\par
}
}
\maketitle

\begin{abstract}
Pretrained cross-modal models, for instance, the most representative CLIP, have recently led to a boom in using pre-trained models for cross-modal zero-shot tasks, considering the generalization properties. However, we analytically discover that CLIP suffers from the text-to-image retrieval hallucination, adversely limiting its capabilities under zero-shot learning: CLIP would select the image with the highest score when asked to figure out which image perfectly matches one given query text among several candidate images even though CLIP knows contents in the image. 
Accordingly, we propose a Balanced Score with Auxiliary Prompts (BSAP)\footnote{Related codes will be released upon acceptance.} to mitigate the CLIP's text-to-image retrieval hallucination under zero-shot learning.
Specifically, we first design auxiliary prompts to provide multiple reference outcomes for every single image retrieval, then the outcomes derived from each retrieved image in conjunction with the target text are normalized to obtain the final similarity, which alleviates hallucinations in the model. 
Additionally, we can merge CLIP's original results and BSAP to obtain a more robust hybrid outcome (BSAP-H).
Extensive experiments on two typical zero-shot learning tasks, \textit{i.e.}, Referring Expression Comprehension (REC) and Referring Image Segmentation (RIS), are conducted to demonstrate the effectiveness of our BSAP. Specifically, when evaluated on the validation dataset of RefCOCO in REC, BSAP increases CLIP's performance by 20.6\%. Further, we validate that our strategy could be applied in other types of pretrained cross-modal models, such as ALBEF and BLIP.

\end{abstract}


\section{Introduction}

\IEEEPARstart{T}{raining} models for specific tasks necessitate extensive data collection, a process that can be remarkably time-consuming and expensive. In contrast, zero-shot learning~\cite{zero-shot-learning01,zero-shot-learning02} eliminates the need for the training step by transferring existing models and making slight modifications to the architecture. Considering its efficiency, zero-shot learning has attracted increasing attention recently.

\begin{figure}[t]
    \centering
    \subfigure[]{
        \includegraphics[width=0.45\textwidth]{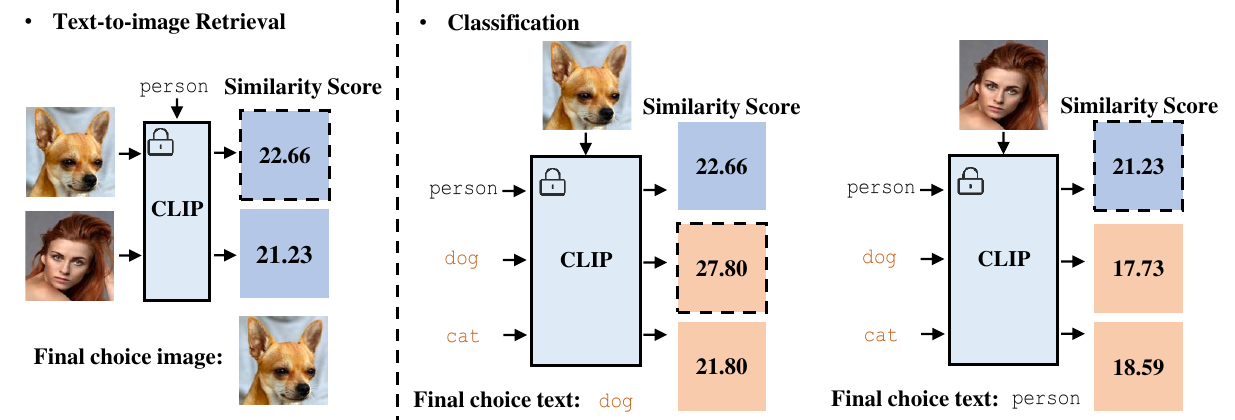}}\\
    \subfigure[]{
        \includegraphics[width=0.45\textwidth]{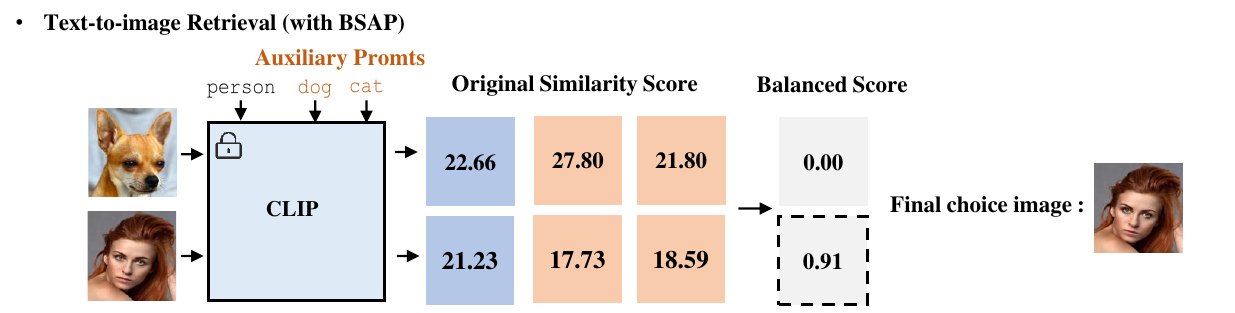}}
    \caption{(a) The illustration of the text-to-image retrieval hallucination of CLIP (ViT-B/32). The text-to-image retrieval hallucination is caused by the range of similarity scores. (b) Performance of the proposed BSAP. When BSAP is employed, the hallucination in the text-to-image retrieval task is resolved.}
    \label{fig:first}
\end{figure}

Recent advancements in foundation models or pretrained models~\cite{tmm01, tmm02} have revolutionized the fields of computer vision and natural language processing and have brought feasible solutions for zero-shot learning considering their capability and generalizability. Among all foundation models, CLIP~\cite{clip} is the most representative one that can align texts and images and is nearly the most commonly used one that improves cross-modal zero-shot tasks, including object detection~\cite{clip_detection}, semantic segmentation~\cite{clip_semantic_segmentation}, image captioning~\cite{clip_caption}, and image-text retrieval~\cite{image_text_retrieval}. 

However, we discovered that CLIP suffers from a type of Text-to-Image Retrieval hallucination, which significantly decreases its zero-shot learning performance. Here, the Text-to-Image Retrieval hallucination refers to the fact that CLIP would obtain a wrong result when asking CLIP to figure out which image perfectly matches one given query text among several candidate images even though CLIP understands the contents in the image. 
We analytically found that in some situations, the CLIP actually can understand the semantic texts and images, and it has been demonstrated that a portion of the hallucinations in CLIP is due to the substantial variance in the range distribution of results computed by CLIP for the same text groups across different images.
For instance, as shown in Fig.~\ref{fig:first} (a), CLIP can identify categories of dog and person in the image. However, CLIP misclassified the dog image as the person text because the similarity score between the dog image and the person text is higher than the similarity score between the person image and the person text. Therefore, we posit that if we can address the issue of imbalanced range distribution in outcomes of multi-modal large foundation models during zero-shot learning, we may be able to mitigate some of the hallucinations in models like CLIP.

In light of the aforementioned analysis, we propose to use Balanced Scores with Auxiliary Prompts (BSAP) for CLIP in text-to-image retrieval, thereby mitigating hallucinations stemming from the imbalanced distribution of the outcome range. 
Specifically, taking the CLIP model as an example, our BSAP initially generates a series of auxiliary result sets for CLIP. %
It then standardizes the scores between each image and given query texts against our auxiliary score sets to achieve a balanced distribution of outcome ranges for the final retrieval. 
As shown in Fig.~\ref{fig:first} (b), with BSAP, the hallucination caused by the imbalanced distribution in CLIP-based method~\cite{reclip} is alleviated.
Note that our auxiliary prompts are pre-defined, and the BSAP requires no training process. In addition, we further obtain hybrid results using BSAP (BSAP-H) that combine balanced results using BSAP with the original results created by Vision-Language Models (VLMs) to obtain more robust final results. We conducted extensive experiments on two typical zero-shot learning tasks, \textit{i.e.}, Referring Expression Comprehension (REC) and Referring Image Segmentation  (RIS). Three datasets, including RefCOCOg~\cite{refcocog01,refcocog02}, RefCOCO+~\cite{refcoco}, and RefCOCO~\cite{refcoco}, and several state-of-the-art CLIP-based methods~\cite{reclip,zero_shot_ris}
are adopted.
The experimental results demonstrate that BSAP improves the CLIP-based methods and some other VLMs. 
Specifically, the results in the REC task of all datasets were increased by an average point of $5.92\%$.

Our contributions are summarized as follows:
\begin{itemize}
 \item We analytically discover that pretrained cross-modal models, for example CLIP, suffer from text-to-image retrieval hallucination, which significantly degrades CLIP's performance in zero-shot learning.

 \item We propose Balanced Scores with Auxiliary Prompts (BSAP) to mitigate hallucinations caused by the imbalanced distribution of results. By calculating a set of reference outcomes, we normalize the original results in conjunction with the reference set to balance the raw results. 

 \item Experimental results on two typical zero-shot learning tasks, namely REC and RIS, demonstrate the capability of our BSAP and BSAP-H. Both tasks have achieved state-of-the-art performance. Additionally, we have briefly tested our BSAP on other vision-language models and the image-to-text retrieval task, and it has also brought about a noticeable improvement.

\end{itemize}

\section{Related Works}

\subsection{Zero-shot Learning}  Zero-shot learning~\cite{zero-shot-learning03,zero-shot-learning04} aims to predict unseen classes by transferring knowledge learned to seen classes. \ %
Recently, CLIP~\cite{clip} and ALIGN~\cite{ALIGN} introduced new perspectives to zero-shot learning through large-scale image-text pre-training. They demonstrate consistent results across various downstream tasks through zero-shot knowledge transfer, including image captioning~\cite{clip_caption}, video action localization~\cite{video_action_localization}, and cross-modal retrieval~\cite{image_text_retrieval}. These studies have directly employed CLIP encoders with minor architectural modifications, eliminating the need for additional training. Our work aligns with this cutting-edge research field.

\subsection{Foundation Models} 
Foundation Models~\cite{foundation_models} are models trained on large corpora of data that, at a very large scale, can generalize to new tasks without any task-specific finetuning. 

The primary method considered in this study is CLIP~\cite{clip}. Our focus on CLIP is rooted in its pre-training on a dataset comprising 400 million image-text pairs gathered from the web, enabling impressive zero-shot image classification performance across various visual domains. CLIP incorporates a dedicated image encoder, either based on the ResNet architecture~\cite{resnet} or a visual transformer~\cite{visual_transformer}, alongside a distinct text transformer. The primary utilization involved the RN50x16 and ViTB/32 versions of CLIP. The limited discriminative capabilities of CLIP are also observed~\cite{clip_limitation01}. 
The paper primarily explores that CLIP the same as MLLMs, also exhibits hallucinations characteristic of such models.

\subsection{Hallucination in Vision-Language Models}

Hallucinations~\cite{Hallucination01,Hallucination02} denote the emergence of text that is irrelevant, factually inaccurate, or nonsensical within a specific context—a significant challenge faced by contemporary large foundational models. This issue may stem from overfitting to particular patterns observed in the training data, a deficiency in grasping real-world factuality, or a failure to contextualize the input provided appropriately~\cite{Hallucination03}. Our research has uncovered that hallucinations are not confined to text-generation tasks. They also manifest in the text-to-image retrieval task domain, which is our study's focus. In this task, discrepancies arise when searching for images that correspond to textual descriptions using VLMs. Nonetheless, the models demonstrate an aptitude for accurately classifying both erroneous and accurate images. This proficiency in classification suggests that VLMs possess the capability to discern between correct and incorrect images. Consequently, our findings have shed light on the phenomenon of hallucinations within the context of large models engaged in text-to-image retrieval tasks.

\subsection{Referring Expression Comprehension and Referring Expression Segmentation}
 Referring expression comprehension (REC)~\cite{rec} and referring expression segmentation (RES)~\cite{ris} require text-to-image retrieval and are typical zero-shot visual grounding~\cite{visual_grounding} tasks. In recent advancements in zero-shot learning for visual grounding, the prevalent approach often employs the transferable CLIP model~\cite{reclip,zero_shot_ris}. The central module involves transforming the task into a text-to-image retrieval task, complemented by additional auxiliary design modules. 
Six methods utilizing CLIP for zero-shot REC and RIS tasks are introduced. For the REC task, Colorful Prompt Tuning (CPT)~\cite{CPT} applies different colors to various proposals and uses a masked language prompt. The pre-trained masked language model~\cite{plm} like CLIP calculates the color with the highest probability. GradCAM~\cite{GradCAM} is a commonly used gradient visualization technique, employed to understand which parts of the input image are most crucial for the model's predictions in computer vision tasks. ReCLIP~\cite{reclip} consists of two components. The first component is a region-scoring method that isolates object proposals through cropping and blurring before passing them to CLIP. The second component is a spatial relation resolver, capable of handling various spatial relationships proposed by the authors. Our approach involved conducting experiments on REC and RES tasks. 

For the RIS task, Region Token~\cite{region_token} is a technique employed in adapting CLIP. Similar to the adaptation process for CLIP, it utilizes region tokens for each mask suggestion in all Transformer layers of the CLIP visual encoder, instead of using superpixels. Finally, it calculates the cosine similarity between the class label of each mask proposal and the text features of CLIP, selecting the mask with the highest score. Cropping~\cite{cropping02} is a commonly used method that leverages CLIP to extract mask or bounding box features in a range of zero-shot dense prediction tasks. Zero-shot-RIS~\cite{zero_shot_ris} introduces a mask-oriented visual encoder capturing both global and local contextual information from the input image. This method can perform fine-grained instance-level groundings segmentation by utilizing instance masks obtained from off-the-shelf mask proposal techniques. Additionally, the author introduces a global-local text encoder, where global features capture complex sentence-level semantics of the entire input expression, while local features focus on target noun phrases extracted by a dependency parser.
\section{Hallucination in CLIP} \label{bias}

\begin{figure}[t]
    \centering
    \subfigure[Examples of misclassification due to hallucinations in Reclip.]{
        \includegraphics[width=0.45\textwidth]{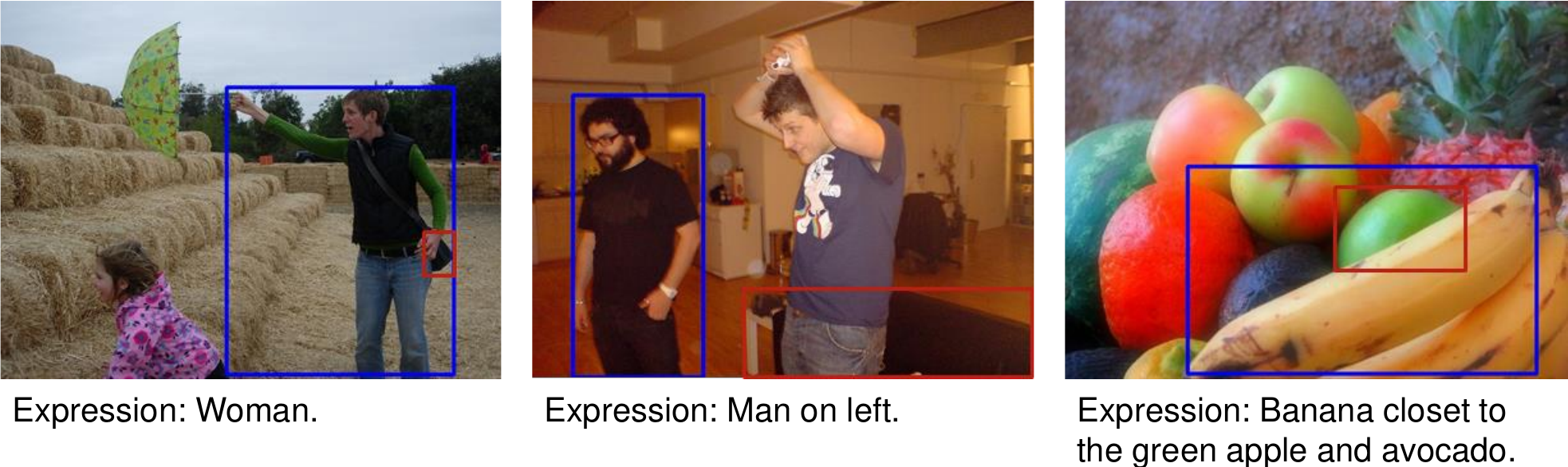}}\\
    \vspace{2pt}
    \rule{\linewidth}{0.75pt}
    \vspace{2pt}
    \subfigure[The imbalanced range of the similarity scores obtained by CLIP.]{
        \includegraphics[width=0.45\textwidth]{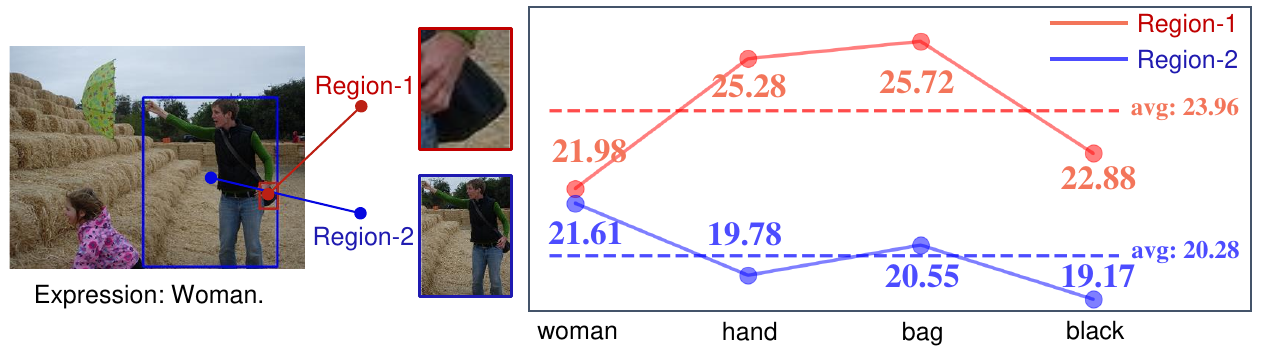}}
    \caption{Examples and analyses of hallucination in CLIP on text-to-image retrieval tasks. The blue bounding boxes represent the Ground Truth (GT), while the red bounding boxes depict the predicted results from ReClip.}
    \label{fig:imbalanced}
\end{figure}

CLIP has the ability to align texts and images, which is frequently used to identify whether one image matches the given text in various zero-shot tasks, such as Image Retrieval with Textual Inversion~\cite{image_text_retrieval}, Referring Expression Comprehension (REC)~\cite{reclip}, and Referring Image Segmentation (RIS)~\cite{zero_shot_ris}. 
The process can be treated as a text-to-image retrieval defined as follows:
\begin{equation}
  r_{\text{pred}}=\arg \max _{m \in \mathcal{U}} \operatorname{Sim}\left(\mathbf{T}_{\text{ref}}, \mathbf{V}_{m}\right),
\end{equation}
where $r_{\text{pred}}$ is the output of predicted result, $\mathcal{U}$ is the set of $n$ images used for retrieval, $\mathbf{T}_{\text{ref}}$ is the textual referring expression, and $\mathbf{V}_{m}$ is the visual feature of the $m$-th image.  $\operatorname{Sim}(\cdot, \cdot)$ is a similarity between image information and text information, where $\operatorname{Sim}(\cdot, \cdot)=\gamma \cdot \operatorname{Cossim}(\cdot, \cdot)$, $\gamma$ is the scaling factor with a fixed value of $100$. 

Hallucination is discovered in CLIP during the text-to-image retrieval task. We observed the issue in the CLIP-base model and proved it use REC tasks in three datasets, \textit{i.e.}, RefCOCOg~\cite{refcocog01,refcocog02}, RefCOCO+~\cite{refcoco}, and RefCOCO~\cite{refcoco} validation, as illustrations. We found that using the Reclip model based on CLIP, nearly $15\%$ of the text cannot be aligned with the correct corresponding images in all the mentioned datasets. We present several examples in Fig.~\ref{fig:imbalanced} (a).
One intuitive reason is that the CLIP misclassifies the semantic contents of these images, leading to the wrong region image selection for the query text. Without training or fine-tuning, it is almost impossible to improve CLIP's distinguishing ability in zero-shot learning.

\begin{figure}[t]
    \centering
    \subfigure[Results for caption ``person".]{\includegraphics[width=0.49\linewidth]{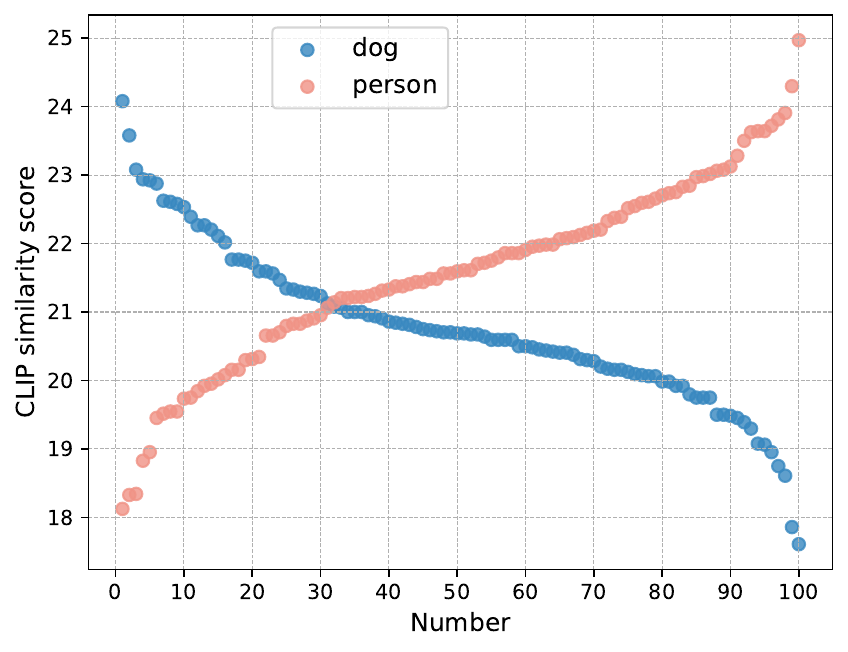}}
    \subfigure[Results for caption ``dog".]{\includegraphics[width=0.49\linewidth]{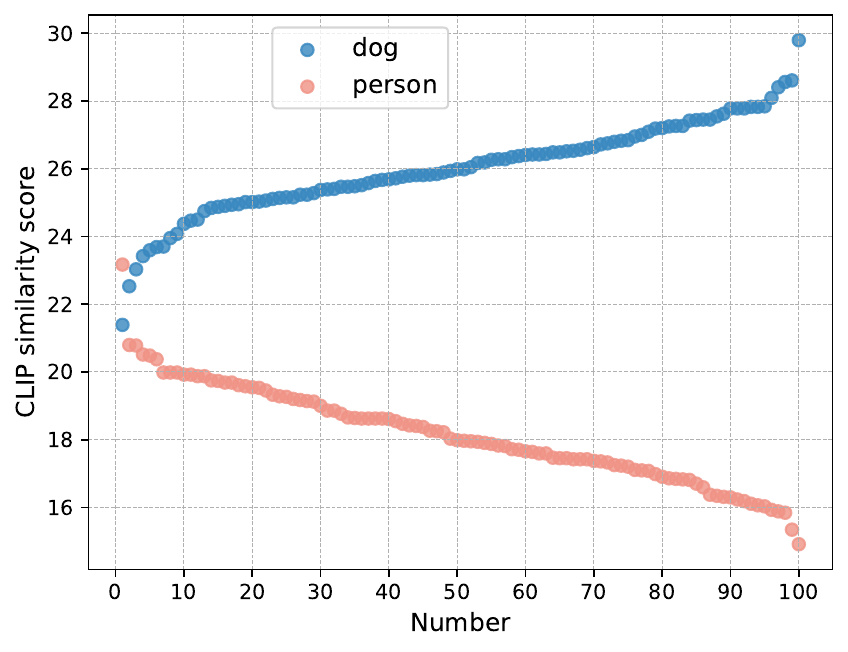}}
\caption{The illustration of the similarity scores generated by the CLIP model, in which 100 images of the dog and 100 images of the human are selected. In the graphs, blue dots represent the result for dog images and red dots represent the result for human images. (a) Similarity score processed by CLIP for the text caption ``person" with 100 images of a dog and 100 images of a human. The results for dog images are sorted from large to small and the results for human images are sorted from small to large. (b) Similarity score processed by CLIP for the caption text ``dog" with 100 images of a dog and 100 images of a human. The results for the dog images are sorted from small to large and the results for human images are sorted from large to small.}
\label{fig:cosine_similarity}
\end{figure}

It was discovered the imbalanced range of the results obtained by the CLIP, which was generally ignored, was one of the reasons that led to hallucination when performing text-to-image retrieval. 
One case is shown in Fig.~\ref{fig:imbalanced} (b).
It was observed that the smallest score of Region-1 is still higher than the highest score of Region-2. Naturally, Region-1 would always be selected. To further analyze the imbalanced range problem, we randomly selected 100 images of dogs and 100 images of persons. To our surprise, when asking whether the images belong to text descriptions of a dog or a person, CLIP can obtain a 99.5\% accuracy. 
This means that the CLIP has a strong ability to understand the contents of dog and person images and text descriptions.
Nevertheless, the Fig.~\ref{fig:cosine_similarity}
shows outcomes between all images and the text descriptions of the dog and person obtained using CLIP.  
The outcomes for 100 dog images paired with the caption ``person" are arranged in descending order, and the outcomes for 100 human images paired with the caption ``person" are arranged in ascending order.  
We observe that the distribution of dog results is mixed with the distribution of person results when using person descriptions with images. 
A significant portion of dog images are higher than those of person images; in other words, dog images will be selected as the match with the caption ``person" in most situations. The imbalanced result ranges lead to hallucinations.

\section{The Proposed Method}

In this section, we design Balanced Scores with Auxiliary Prompts (BSAP) to alleviate the hallucinations for CLIP in zero-shot learning. Firstly, we present the details to obtain balanced results and then give an explanation of how to generate auxiliary prompts under zero-shot learning. In addition, we introduce hybrid results using BSAP (BSAP-H), which combines our BSAP and original outcomes. 

\subsection{Balanced Scores with Auxiliary Prompts} \label{BASI}
\begin{figure}[]
\centering
\includegraphics[width=0.8\linewidth]{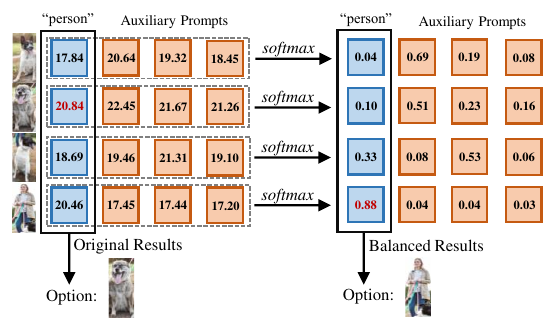}
\caption{The process of BSAP to get balanced results with auxiliary prompts.}
\label{fig:similarity}
\end{figure}

In light of the analysis in Section~\ref{bias},
the diverse ranges of results would lead to hallucinations. Intuitively, if we can balance all outcomes in the same space, the problem could be mitigated. Inspired by the normalization strategies~\cite{norm01,norm02}, we propose Balanced Scores with Auxiliary Prompts (BSAP).

As shown in Fig.~\ref{fig:similarity}, there are several auxiliary text prompts that could be used as calibration. Similarity scores can be adjusted with these auxiliary prompts. Specifically, for $m$-th image with one textual referring expression and auxiliary prompts, similarity scores of the image are calculated as follows:
\begin{equation}
\text{Sim}_{r}=\operatorname{Sim}\left(\mathbf{T}_{\text{ref}}, \mathbf{V}_{m}\right),
\end{equation}
\begin{equation}
\text{Sim}_{\text{ap}_i}=\operatorname{Sim}\left(\mathbf{T}_{\text{ap}_i}, \mathbf{V}_{m}\right),
\end{equation}
where $\text{Sim}_*$ refers to similarity scores of CLIP, $\mathbf{V}_{m}$ refers to the visual feature of the $m$-th image, $\mathbf{T}_{\text{ref}}$ refers to textual embeddings of query text, and $\mathbf{T}_{\text{ap}_i}$ refers to textual embeddings of auxiliary prompts, $i\in [A]$ and $A$ is the number of auxiliary prompts.

Then, we normalize scores and obtain balanced similarity scores. The process is defined as follows:
\begin{equation}
\mathbf{BS}=\frac{e^{\text{Sim}_r}}{e^{\text{Sim}_r}+e^{\sum_{i\in A}\text{Sim}_{\text{ap}_i}}},
\end{equation}
where $\mathbf{BS}$ is final balanced similarity scores. The process is similar to a SoftMax function. 
Note that other normalization strategies can also be adopted, such as feature scaling. 
We select SoftMax because it experimentally obtains sufficient improvements. If the auxiliary prompts are sufficient, we can obtain reliable balanced scores, since the normalization would balance the range of similarity scores of different objects in a similar space. The final prediction is determined as follows:
\begin{equation}
r_{\text{pred}}^\text{BS}=\arg \max _{m \in \mathcal{U}}{\mathbf{BS}_m},
\end{equation}
where $\text{BS}_m$ refers to the balanced result score of $m$-th image.

\subsection{Auxiliary Prompts Generation} \label{additional}

One key component of our balanced outcome scores is how to obtain the auxiliary prompts under zero-shot learning settings since there are no reference templates. We design our auxiliary prompts from two aspects: contents and templates. 

\paragraph{Contents} Most methods use CLIP to align the texts and images. The texts would contain at least one object. Since we cannot know what would meet in zero-shot learning settings, we select object categories in the existing datasets as contents in the auxiliary prompts, as existing datasets contain most of the commonly seen objects. We have tested the object categories in COCO~\cite{COCO}, CIFAR~\cite{CIFAR}, and Caltch~\cite{CIFAR} and experimentally found that categories combined from COCO and CIFAR obtained better results. 

\paragraph{Templates} Generally, the original prompt of CLIP not only contains an individual object but a fixed template, such as ``a photo of". We also adopt the templates to generate our auxiliary prompts. The formula for auxiliary prompts is defined as follows:
\begin{equation}
 t_{\text {add }}(i)=t_{\text {template }}+t_{\text {head }}(i), i \leq A,
\end{equation}
where $t_{\text {template }}$ is a fixed template and $t_{\text {head }}$ is object categories we selected for content. The query text will also be added with the fixed template. 

Moreover, we find that in several tasks~\cite{CoOp,CoCoOp}, the query texts are sentences with objects, such as ``piece of broccoli at far right". %
We experimentally discovered that the template length would also affect the performance of CLIP. Consequently, we use ChatGPT~\cite{GPT3} to adjust the fixed templates with the same length of the input query text. The prompt for ChatGPT to adjust is defined as follows: Provide me with a template of the number of words is $X$ similar to ``a photo of $<>$". Note that, in such a situation, the query text will not be added with the fixed template.

\subsection{Hybrid Scores Using BSAP}\label{HYSI}

We drew inspiration from~\cite{hybird} to adopt hybrid scores that combined our balanced scores using BSAP and original similarity scores of CLIP to obtain more robust retrieval results. The hybrid scores formula is as follows:
\begin{equation}\label{alpha}
r_{\text {pred}}^{\text{HS}}=\arg \max _{m \in \mathcal{U}} \left({\alpha} _m \mathbf{BS}_m + (1-\alpha) \mathbf{S}_{\text{ref}} \right),
\end{equation}
where $\alpha $ is a constant parameter, $\mathbf{BS}_m$ is the final balanced score, and $\mathbf{S}_{\text{ref}_m}$ represents the original scores considering only the textual referring expression and all the image regions:
\begin{equation}
  \mathbf{S}_{\text{ref}_m}={\operatorname{softmax}} (\operatorname{Sim}\left(\mathbf{T}_{\text{ref}}, \mathbf{V}_{m}\right)).
\end{equation}

Hybrid scores incorporate both the original scores derived from focused query text and balanced scores using our auxiliary prompts which can alleviate hallucinations in text-to-image retrieval, thereby achieving more stable results.

\section{Examples of Applying BSAP to different zero-shot models}

\subsection{Applying BSAP to Reclip in REC Task}

\begin{figure*}[ht]
\centering
\includegraphics[width=\linewidth]{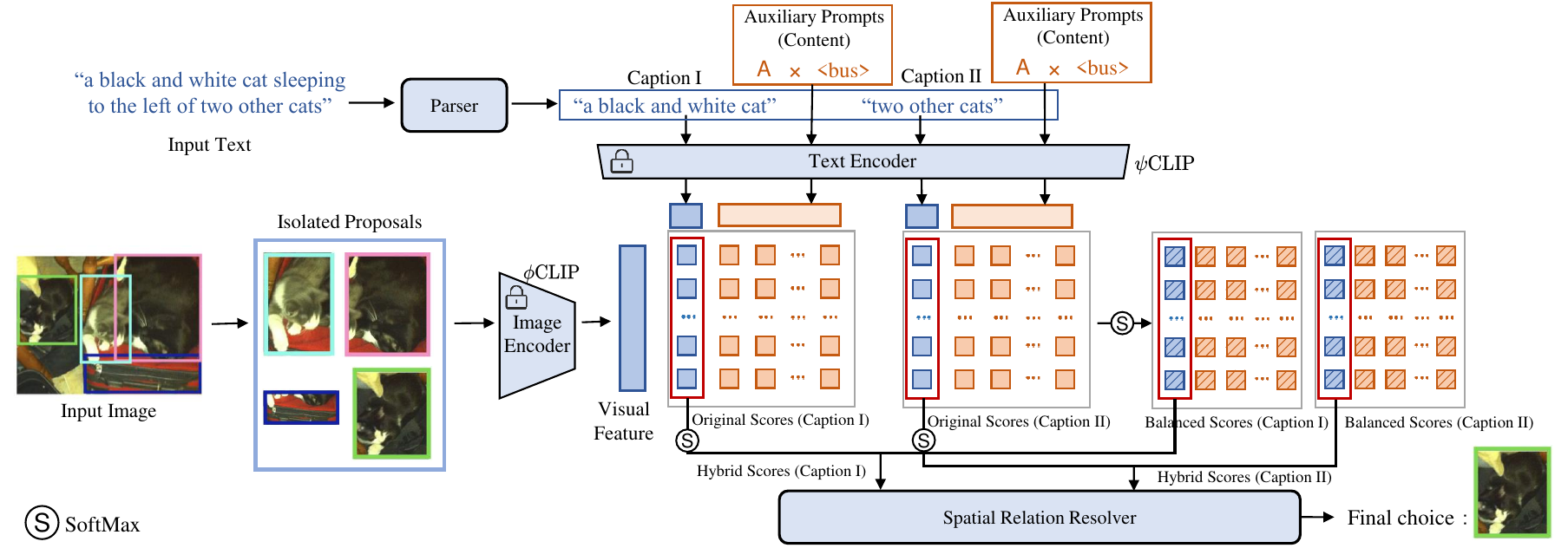}
\caption{Apply BSAP to Reclip. Blue sections represent the original core of ReCLIP. Orange sections denote the components that our BSAP and BSAP-H integrate into ReCLIP. Red boxes highlight original scores and our BSAP.}
\label{fig:BSAP_Reclip}
\end{figure*}

The ReCLIP consists of two key components in addition to CLIP. One mechanism utilizes CLIP to score object proposals, while the other involves a method for handling spatial relationships between objects.
\begin{itemize}
    \item \textbf{Isolated Proposal Scoring (IPS)}: The IPS technique involves isolating individual proposals through image cropping and blurring. Subsequently, these isolated proposals are scored based on a given expression using CLIP.
    \item \textbf{Spatial Relation Resolver}: In order to address spatial relations, they proposed spatial heuristics. This involves decomposing an expression into subqueries. CLIP is then leveraged to compute proposal probabilities for each subquery. The outputs of all subqueries are amalgamated using simple rules. This spatial heuristic mechanism is designed to enhance the handling of spatial relationships within the ReCLIP framework.
\end{itemize}
As shown in Fig.~\ref{fig:BSAP_Reclip}, the blue section represents the original core of the ReCLIP model. The orange section denotes the components that our BSAP and BSAP-H integrate into ReCLIP. The red boxes highlight the original similarity scores and our BSAP. In the figure, the parser divides the sentence into multiple segments. Each caption contains a core word. The number of divisions corresponds to the number of times CLIP is used, and accordingly, auxiliary prompts are applied the same number of times. The hybrid scores (BSAP-H) serve as the input to the Spatial Relation Resolver, replacing the original IPS results.

\begin{figure*}[ht]
\centering
\includegraphics[width=\linewidth]{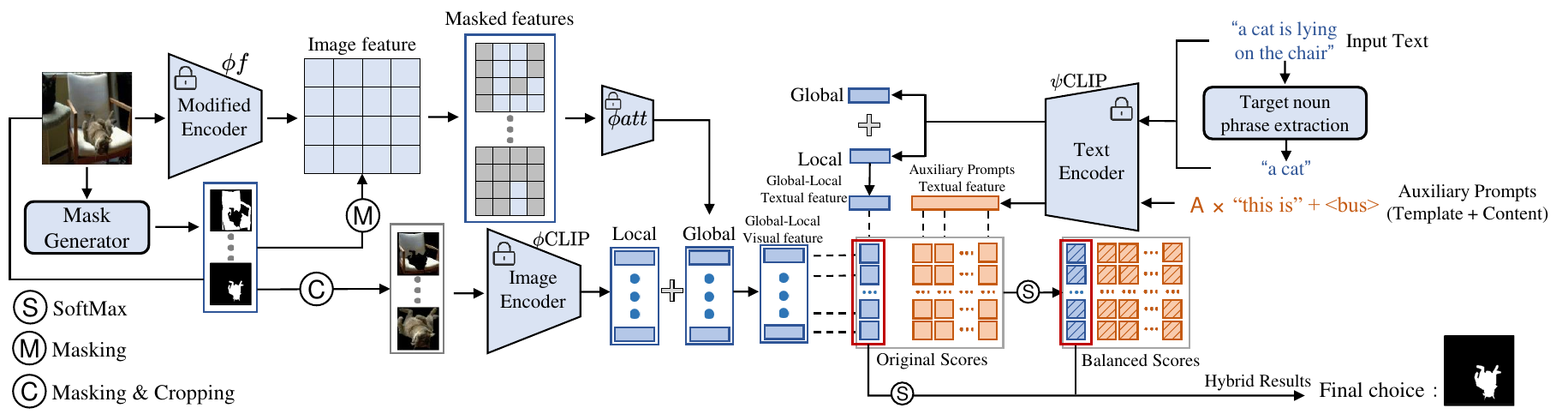}
\caption{Apply BSAP to Zero-shot-RIS. Blue sections represent the original core of Zero-shot-RIS. Orange sections denote components that our BSAP and BSAP-H integrate into Zero-shot-RIS. Red boxes highlight original scores and BSAP.}
\label{fig:BSAP_ris}
\end{figure*}

\subsection{Apply BSAP to Zero-shot-RIS in RIS Task}
Zero-shot-RIS captures local information in images as input for the image encoder of CLIP and captures the fusion of local key terms from the query text and overall context as input for the text encoder of CLIP. The two core components of Zero-shot-RIS are as follows:
\begin{itemize}
    \item \textbf{Mask-Guided Visual Encoder}: Introduce a mask-guided visual encoder that effectively captures both global and local context information within an image, guided by a given mask. This encoder serves to understand the spatial relationships and features associated with the target object in the image.
    \item \textbf{Global-Local Textual Encoder}: Develop a global-local textual encoder that considers both the local and global context of the referring expression. The local context is captured through a target noun phrase, providing specificity to the description, while the global context is encompassed by the entire sentence of the expression.
\end{itemize}
Finally, establishing a mechanism for combining features from two distinct context levels: visual and textual. By integrating information from both local and global perspectives, the method aims to achieve a comprehensive understanding of the scene and a specific characterization of the target object.

As shown in Fig.~\ref{fig:BSAP_ris}, the blue section represents the original core of the Zero-shot-RIS model. The orange section denotes the components that our BSAP and BSAP-H integrate into Zero-shot-RIS. The red boxes highlight the original similarity scores and BSAP.

\section{Experiment}

We conducted extensive experiments to demonstrate the capability of our BSAP. Four questions have been answered in this section:

\begin{itemize}
 \item Q1: Can our BSAP improve the CLIP in zero-shot learning tasks? 
\item Q2: Can our BSAP truly mitigate the text-to-image retrieval hallucinations for CLIP?
\item Q3: What is the contribution of each component in BSAP?
\item Q4: Can our BSAP mitigate the hallucinations and improve the performance of other VLMs?
\end{itemize}

Additionally, we investigated whether BSAP could alleviate hallucinations in the image-to-text retrieval task.

\subsection{Experimental Settings}
We first introduce the experimental settings. Two typical zero-shot learning tasks are selected as the test beds: REC and RIS. The datasets, evaluation metrics, and implementation details are given subsequently.

\subsubsection{Datasets}
In both tasks, RefCOCO, RefCOCO+, and RefCOCOg datasets are utilized. 
RefCOCO, RefCOCO+, and RefCOCOg datasets consist of 19,994, 19,992, and 26,711 images, respectively, with 142,210, 141,564, and 104,560 images containing textual referring expressions. RefCOCO and RefCOCO+ expressions are relatively shorter, with 3.61 and 3.53 words per expression on average, while RefCOCOg sentences are longer and more complex, around 8.43 words per expression on average.  

\subsubsection{Evaluation Metrics}
The measurement metrics in our experiments follow the Reclip model~\cite{reclip} and the Zero-shot-RIS~\cite{zero_shot_ris} model for REC and RIS, respectively. 

\noindent\textbf{The Reclip model} for the REC task uses the accuracy (acc)~\cite{reclip}. When the Intersection over Union (IoU) between the predicted target and the actual target exceeds 0.5, it is deemed correct. Conversely, if it falls below 0.5, it is considered incorrect.

\noindent\textbf{The Zero-shot-RIS model} for the RIS task uses the overall Intersection over Union (oIoU) and the mean Intersection over Union (mIoU).  oIoU is calculated by dividing the total area of intersection by the total area of the union, where the total area is computed by aggregating over all examples. Additionally, mIoUs calculates the average IoU across all examples while taking object sizes into consideration, as reported in~\cite{miou01,miou02}.

\subsubsection{Implementation Details}
The core $t_{\text {head }}$ are selected for auxiliary prompts including 80 classes of COCO~\cite{COCO} and 100 classes of CIFAR~\cite{CIFAR}. ``a photo of" is consistently used as the template. 
Regarding the template length, the Reclip model undergoes further analysis and processing of the text, leading to variable lengths. 
For the Zero-shot-RIS model, the corresponding input remains unprocessed. Templates with a length of 2 are chosen for  RefCOCO and RefCOCO+ datasets, and a length of 5 for RefCOCOg dataset. 
Generated templates of different lengths are all generated based on ChatGPT~\cite{GPT3}. As for the constant parameter $\alpha $, $\alpha = 0.75$ is selected for the REC task and $\alpha = 0.5$ is selected for the RIS task. Since it is a zero-shot task, the constant parameter $\alpha$ in different tasks is only tested on the val data set of RefCOCOg. 
Better parameter values may exist in other data sets. Our strategy and the enhanced models were operated in an environment with one RTX-2080 GPU, requiring no additional training.

\subsection{Improvements of CLIP with BSAP (Q1)}
 
\subsubsection{Performance on zero-shot REC}
\begin{table*}[]
\small
\centering
\caption{Accuracy on RefCOCOg, RefCOCO+, and RefCOCO in zero-shot REC. The best results in each column are in \textbf{bold}, and previously best results are \underline{underlined}. Reclip adopts  RN50x16 and ViT-B/32 CLIP models. CPT-adapted is an adapted version of CPT-Blk. Supervised model UNINEXT-H~\cite{uniter}. All methods except UNITER-L adopt GT proposals from official annotations of MSCOCO as candidates. When using other VLMs, they all integrated into Reclip, replacing the original CLIP model. ALBEF uses the 14M version, while BLIP uses the ViT-L version.}
{
\begin{tabular}{lcccccccc}
\toprule
\multirow{2}[2]{*}{{\bf Model}} & \multicolumn{2}{c}{RefCOCOg} & \multicolumn{3}{c}{RefCOCO+} & \multicolumn{3}{c}{RefCOCO} \\
& \multicolumn{1}{c}{\textbf{Val}} &  \multicolumn{1}{c}{\textbf{Test}} &  \multicolumn{1}{c}{\textbf{Val}} &  \multicolumn{1}{c}{\textbf{TestA}} &  \multicolumn{1}{c}{\textbf{TestB}}  & \multicolumn{1}{c}{\textbf{Val}} &  \multicolumn{1}{c}{\textbf{TestA}} &  \multicolumn{1}{c}{\textbf{TestB}}\\
\midrule
Random & 20.18 & 20.34 & 16.73 & 12.57 & 22.13 & 16.37 & 12.45 & 21.32 \\
\midrule
UNITER-L~\cite{uniter} & 87.85 & 87.73 & 84.25 & 86.34 & 79.75 & 91.84 & 92.65 & 92.65 \\
\midrule
\textbf{CLIP-based Methods} & & & & & & & & \\
CPT-adapted~\cite{CPT} & 25.96 & 25.87 & 25.44 & 22.00 & 28.74 & 24.79 & 21.6 & 28.89 \\
GradCAM~\cite{GradCAM} & 56.82 & 56.15 & 51.10 & \underline{57.79} & 43.24 & 46.68 & \underline{51.99} & 40.10 \\

ReCLIP w/o relations~\cite{reclip} & 65.32 & 65.59 & 51.54 & 51.80 & 50.85 & 45.66 & 45.13 & 45.40 \\
ReCLIP~\cite{reclip}  & \underline{68.08} & \underline{67.05} & \underline{52.12} & 51.61 & \underline{52.03} & \underline{50.51} & 47.11 & \underline{54.94} \\ 
\midrule

ReCLIP + BSAP  & 67.73 & 68.33 & \textbf{59.80} & \textbf{65.40} & 52.96 & \textbf{60.90} & \textbf{64.72} & \textbf{58.47} \\
ReCLIP + BSAP-H & \textbf{69.65} & \textbf{68.93} & 57.23 & 60.37 & \textbf{54.00} & 57.57 & 58.58 & 58.27 \\

\midrule
\textbf{Other VLMs} & & & & & & & & \\
ALBEF~\cite{albef} & 72.23 & 72.45 & 63.78 & 65.45 & 61.45 & 63.57 & 65.49 & 61.79 \\
ALBEF~\cite{albef} + BSAP & 72.98 & 73.06 & 66.24 & 68.78 & 64.49 & 65.79 & 68.99 & 63.96 \\
BLIP~\cite{blip} & 73.13 & 72.95 & 64.78 & 67.29 & 62.42 & 65.09 & 67.78 & 64.32 \\
BLIP~\cite{blip} + BSAP & 74.54 & 74.14 & 67.67 & 68.78 & 64.55 & 65.79 & 69.45 & 65.94 \\
\bottomrule
\end{tabular}
}
\label{tab:REC_Result}
\end{table*}

In Table~\ref{tab:REC_Result}, the performance of Reclip~\cite{reclip} with BSAP and BSAP-H is presented in referring expression comprehension (REC), compared to CPT-adapted~\cite{CPT}, GradCAM~\cite{GradCAM}, Reclip~\cite{reclip} on RefCOCO, RefCOCO+, and RefCOCO datasets. To avoid the influence of accuracy variations in proposals from different detection models, Ground-truth proposals of MSCOCO %
are utilized as inputs following Reclip.

When only employing the BSAP, there is a significant improvement observed on the RefCOCO+ and RefCOCO datasets, with the highest boost reaching $12.73\%$. BSAP-H also consistently outperforms the original baseline methods across the board. The results demonstrate that our BSAP has the ability to improve the CLIP on REC. 

\subsubsection{Performance on Zero-shot RIS}

\label{sec:RIS_Result}

\begin{table*}[]
\small
\centering

\caption{Comparison with Zero-shot RIS baseline methods on three standard benchmark datasets in zero-shot RIS task. All baseline methods use FreeSOLO as the mask proposal network. The best zero-shot results in each column are in \textbf{bold}, and the previously best results are \underline{underlined}. FreeSOLO upper-bound is computed between the GT mask and the maximum overlapped FreeSOLO mask with the GT mask. When using other VLMs, they all integrated into Reclip, replacing the original CLIP model. ALBEF uses the 14M version, while BLIP adopts the ViT-L version.}

\resizebox{1.0\textwidth}{!}{
\begin{tabular}{c|ll|cc|ccc|cccc}
\hline
\multirow{2}{*}{Metric}  & \multirow{2}{*}{Methods}            & \multirow{2}{*}{\begin{tabular}[c]{@{}l@{}}Visual\\ Encoder\end{tabular}}  & \multicolumn{2}{c|}{RefCOCOg}  & \multicolumn{3}{c|}{RefCOCO+} & \multicolumn{3}{c}{RefCOCO}  \\ \cline{4-12} 
& &   & Val & Test & Val & TestA  & TestB & Val & TestA  & TestB  \\ \hline
\multirow{12}{*}{oIoU} & Supervised  method~\cite{lavt} &   
& 61.24 & 62.09 & 62.14 & 68.38 & 55.10 & 72.73 & 75.82 & 68.79 \\ \cline{2-12}
& \textbf{Zero-Shot Baselines} &  &   &  &  &  &  &  &  &  \\
& Region token~\cite{region_token} & ViT-B/32
& 25.52  & 25.38 & 22.61 & 20.91 & 23.46 & 21.71 & 20.31 & 22.63 \\
& Cropping & ResNet-50                                                     
& 28.20 & 27.64 & 23.95 & 22.03 & 23.49 & 22.36 & 20.49 & 22.69 \\
& Cropping & ViT-B/32
& 28.69 & 27.51 & 24.09 & 22.42 & 23.93 & 22.73 & 21.11 & 23.08 \\ 
& Zero-shot-RIS~\cite{zero_shot_ris} & ResNet-50    
& 30.07 & 29.83 & 25.87 & 24.61 & 25.61 & 24.58 & 23.38 & 24.35 \\
& Zero-shot-RIS & ViT-B/32 
& \underline{31.11} & \underline{30.96} & \underline{26.16} & \underline{24.90} & \underline{25.83} & \underline{24.88} & \underline{23.61} & \underline{24.66}\\  
\cdashline{2-12}[1pt/1pt]
& Zero-shot-RIS + BSAP & ResNet-50    
& 30.24 & 30.14 & 26.39 & 25.51 & \textbf{25.95} & 24.91 & 24.28 & 24.62  \\
& Zero-shot-RIS + BSAP & ViT-B/32 
& 31.07 & 31.07 & 26.14 & \textbf{25.72} & 25.39 & 25.05 & 24.80 & 24.55  \\
& Zero-shot-RIS + BSAP-H & ResNet-50 
& 30.14 & 29.91 & 26.28 & 25.10 & 25.71 & 24.89 & 24.01 & 24.61 \\
& Zero-shot-RIS + BSAP-H & ViT-B/32                                            
& \textbf{31.45} & \textbf{31.31} & \textbf{26.43} & 25.59 & 25.89 & \textbf{25.16} & \textbf{24.89} & \textbf{24.70} \\
\cdashline{2-12}[1pt/1pt]

& FreeSOLO upper-bound & \multicolumn{1}{c|}{-}
& 42.08 & 42.52 & 43.52 & 42.17 & 42.52 & 43.80 & 48.81 & 48.96 \\ 

\cline{2-12}

& \textbf{Other LLMs}   &  &  &  &  &  &  &  &  &  \\
& ALBEF(14M)~\cite{albef} & ViT-B/16                                         & 31.49 & 31.25 & 27.45 & 26.49 & 26.89 & 25.46 & 25.41 & 25.43 \\
& ALBEF(14M) + BSAP  & ViT-B/16                                 & 31.65 & 31.36 & 27.69 & 26.54 & 27.10 & 25.59 & 25.56 & 25.47 \\ 
& BLIP~\cite{blip} & ViT-L/16                                               & 31.54 & 31.29 & 27.49 & 26.78 & 26.87 & 25.56 & 25.74 & 25.56 \\
& BLIP + BSAP  & ViT-L/16                                       & 31.67 & 31.40 & 27.65 & 26.80 & 26.97 & 25.64 & 25.91 & 25.64 \\

\hline
\hline
\multirow{12}{*}{mIoU} & \textbf{Zero-Shot Baselines} &  &  &  &  &  &  &  &  &  \\
& Region token~\cite{region_token} & ViT-B/32                 
& 27.57 & 27.34 & 24.51 & 22.64 & 25.37 & 23.43 & 22.07 & 24.62 \\
& Cropping & ResNet-50 
& 31.27 & 30.87 & 26.31 & 23.94 & 25.69 & 24.31 & 22.37 & 24.66 \\
& Cropping & ViT-B/32
& 31.88 & 30.94 & 26.33 & 24.06 & 26.46 & 24.83 & 22.58 & 25.72 \\ 
& Zero-shot-RIS~\cite{zero_shot_ris} & ResNet-50
& 33.02 & 33.12 & \underline{28.22} & \underline{26.54} & \underline{27.86} & \underline{26.70} & \underline{24.99} & 26.48 \\
& Zero-shot-RIS & ViT-B/32       
& \underline{33.52} & \underline{33.67} & 27.80 & 25.64 & 27.84  & 26.20 & 24.94 & \underline{26.56} \\  
\cdashline{2-12}[1pt/1pt]
& Zero-shot-RIS + BSAP & ResNet-50                                                            & 33.23 & 33.31 & \textbf{29.24} & 28.32 & \textbf{28.78} & \textbf{27.75} & 26.86 & \textbf{27.47}  \\
& Zero-shot-RIS + BSAP & ViT-B/32  
& 34.31 & \textbf{34.56} & 28.86 & \textbf{28.39} & 28.35 & 27.59 & \textbf{27.28} & 27.28  \\
& Zero-shot-RIS + BSAP-H & ResNet-50 
& 33.28 & 33.12 & 28.86 & 27.70 & 28.50 & 27.34 & 26.36 & 27.31 \\
& Zero-shot-RIS + BSAP-H & ViT-B/32 
& \textbf{34.52} & 34.48 & 28.67 & 27.80 & 28.30 & 27.34 & 27.02 & 27.10 \\
\cdashline{2-12}[1pt/1pt]
& FreeSOLO upper-bound~\cite{freesolo} & \multicolumn{1}{c|}{-}       & 48.25 & 46.62 & 50.43 & 48.28 & 46.62 & 50.62 & 52.44 & 52.91 \\ 

\cline{2-12}

& \textbf{Other VLMs}   &  &  &  &  &  &  &  &  &  \\
& ALBEF(14M)~\cite{albef} & ViT-B/16                                         & 35.14 & 35.27 & 29.21 & 30.14 & 29.79 & 28.78 & 27.49 & 27.79 \\
& ALBEF(14M) + BSAP  & ViT-B/16                                 & 36.65 & 35.56 & 29.78 & 30.56 & 30.45 & 29.30 & 28.20 & 28.24 \\ 
& BLIP~\cite{blip} & ViT-L/16                                               & 36.49 & 36.75 & 30.46 & 30.04 & 30.49 & 29.02 & 29.14 & 28.45 \\
& BLIP + BSAP  & ViT-L/16                                       & 37.03 & 37.25 & 30.99 & 30.63 & 30.97 & 29.61 & 29.46 & 29.00 \\ 

\cline{2-12}
& \textbf{Weakly-supervised method}   &  &  &  &  &  &  &  &  &  \\
& TSEG~\cite{tseg} & ViT-S/16$^\dagger$                                                        
& 25.95 & - & 22.62 & - & - & 23.41 & - & - \\ \hline
\end{tabular}
\label{tab:RIS_Result}
}
\end{table*}

In Table~\ref{tab:RIS_Result}, the performance of Zero-shot-RIS with our methods in the RIS task is presented, employing both BSAP and BSAP-H, along with other baselines on RefCOCO, RefCOCO+, and RefCOCO datasets in terms of oIoU and mIoU metrics. For a fair comparison, FreeSOLO~\cite{freesolo} mask proposals are used to generate the final output masks for all methods, including Region token~\cite{region_token}, Cropping, and Zero-shot-RIS~\cite{zero_shot_ris}.

The experimental results indicate that our method outperforms Zero-shot-RIS and other baseline methods comprehensively. The original model with BSAP outperforms previous methods across all datasets in terms of mIoU metrics, while the model with BSAP-H excels previous methods in all metrics across all datasets. Although there is still a gap compared to fully supervised reference image segmentation methods~\cite{lavt}, our approaches significantly improve performance compared to baselines with similar upper bounds. Additionally, Zero-shot-RIS with BSAP-H outperforms supervised method~\cite{tseg}.

Our BSAP can be regarded as a plug-in module to be easily applied in CLIP. The improvements in REC and RIS on two state-of-the-art CLIP-based methods also validate the generalizability of our BSAP.

\begin{table}[]
\centering
\caption{BSAP enhances CLIP's discriminative ability in the REC task, observed in the RefCOCOg, RefCOCO+, and RefCOCO validation datasets. The proportion of the predicted object with the same category as the ground truth.}
\label{tab:outside}
\centering
\renewcommand\arraystretch{1.1}
\setlength\tabcolsep{7pt}
\resizebox{0.38\textwidth}{!}{
\begin{tabular}{c|cc}
\hline 
Dataset & Reclip & Reclip + BSAP\\
\hline
RefCOCOg & $84.805$ & $86.441$  ($+1.636$) \\
RefCOCO+ & $84.672$ & $89.699$  ($+5.027$) \\
RefCOCO  & $84.475$ & $89.843$  ($+5.368$) \\
\hline
\end{tabular}
}
\end{table}

\begin{table}[]
\centering
\caption{BSAP enhances CLIP's discriminative ability in the REC task, observed in the RefCOCOg, RefCOCO+, and RefCOCO validation datasets. The proportion of cases where the predicted object and the ground truth object are the same when their categories same.}
\label{tab:inside}
\renewcommand\arraystretch{1.1}
\setlength\tabcolsep{7pt}
\resizebox{0.38\textwidth}{!}{
\begin{tabular}{c|cc}
\hline 
Dataset & Reclip & Reclip + BSAP\\
\hline
RefCOCOg & $77.851$ & $80.575$  ($+2.724$) \\
RefCOCO+ & $61.449$ & $72.911$  ($+11.412$) \\
RefCOCO  & $55.768$ & $67.785$  ($+12.017$) \\
\hline
\end{tabular}
}
\end{table}

\begin{figure}[]
\centering
\includegraphics[width=0.8\linewidth]{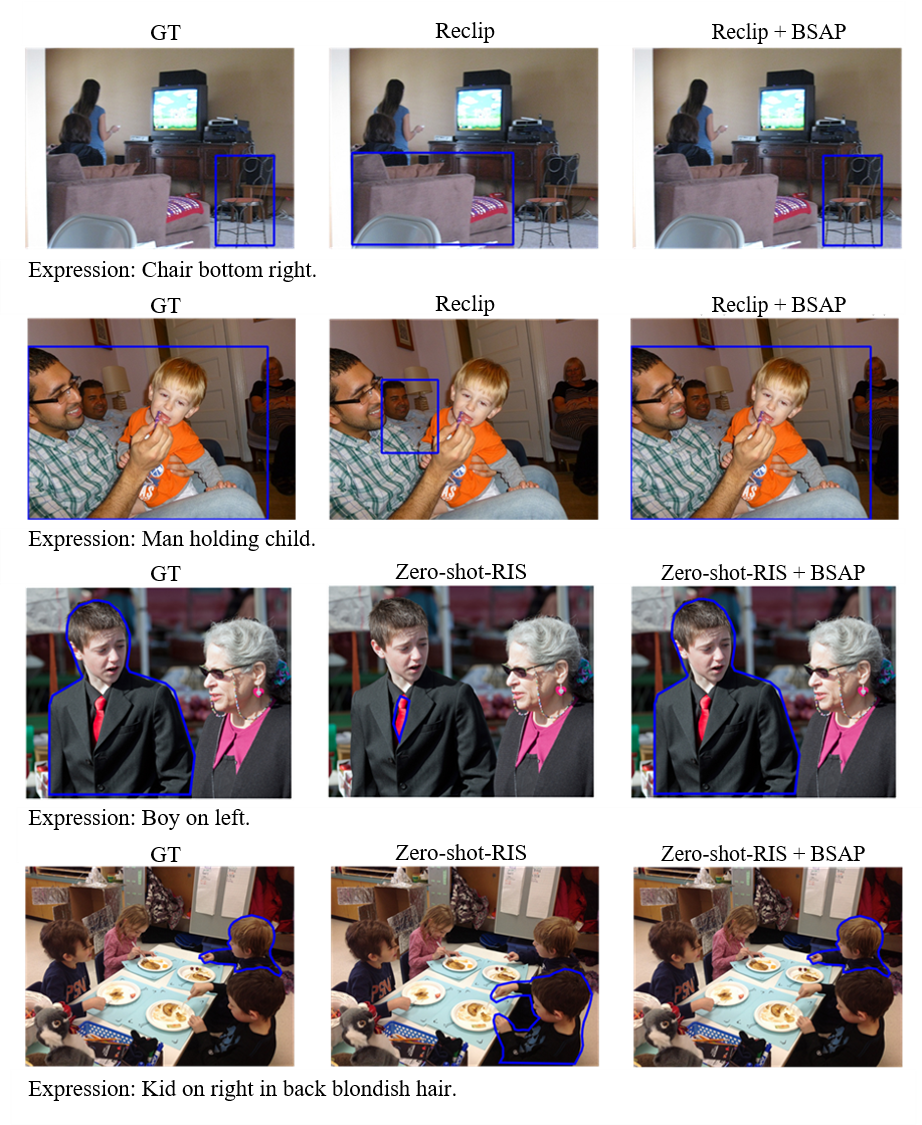}
\caption{Examples of the improvement in CLIP's recognition capability using the hybrid similarity. For the REC task, GT is utilized from the COCO dataset as proposal bounding boxes. In the case of the RIS task, GT is utilized from the COCO dataset as mask proposals.}
\label{fig:result_vis}
\end{figure}

\subsection{On Mitigating Hallucinations in Text-to-Image Retrieval Tasks Using BSAP (Q2)}
As aforementioned analysis, the text-to-image retrieval hallucination of CLIP can be addressed by balancing the original scores with auxiliary prompts. To demonstrate the effectiveness more clearly, we use GT (Ground Truth) from MSCOCO instances as our proposal boxes and mask proposals. We adopt Reclip in REC as the test method.
Table~\ref{tab:outside} and Table~\ref{tab:inside} present the extent to which the text-query balanced similarity enhances the recognition accuracy of the CLIP model in the REC task. Specifically, in Table~\ref{tab:outside}, the results are the accuracy of the query texts being correctly matched with the image regions that contain the same categories. 
In Table~\ref{tab:inside}, the results are the accuracy that the query texts are correctly matched with the image regions that contain not only the same categories but also the correct attribute description. 

The recognition ability between different object categories in Table~\ref{tab:outside} has improved by an average of $4\%$, and the fine-grained recognition ability between objects of the same category in Table~\ref{tab:inside} has improved by an average of $9\%$. The improvements demonstrate that our BSAP could mitigate CLIP's text-to-image retrieval hallucination and enable CLIP to obtain better retrieval performance.

As shown in Fig.~\ref{fig:result_vis}, to provide a more intuitive illustration, we present the visual representations using Reclip and Zero-shot-RIS. It can be seen that with BSAP, the CLIP-based methods can locate the correct image regions better.

\subsection{Ablation Study (Q3)}

\begin{table}[]
\footnotesize
\centering
\caption{In validation datasets of RefCOCOg, RefCOCO+, and RefCOCO, we studied the effect of different dataset categories in contents of our BSAP on accuracy for the REC.}
\renewcommand\arraystretch{1.1}
\setlength\tabcolsep{2pt}
\resizebox{0.47\textwidth}{!}{
\begin{tabular}{ccc|c|c|c}
\hline
\multicolumn{3}{c|}{Dataset (Number of categories)} & \multirow{2}{*}{RefCOCOg} & \multirow{2}{*}{RefCOCO+} & \multirow{2}{*}{RefCOCO} \\
\cline{1-3}
COCO (80)  & CIFAR (100) & Caltech (101)& &  &  \\ \hline
$\checkmark$ & - & - & 66.524 & 58.961 & 60.246   \\
- & $\checkmark$ & - & 66.156 & 59.246 & 59.978   \\
- & - & $\checkmark$ & 65.067 & 57.489 & 58.745   \\
$\checkmark$ & $\checkmark$ & -  & \textbf{66.728} & \textbf{59.342} & \textbf{60.495}  \\
$\checkmark$ & $\checkmark$ & $\checkmark$  & 65.728 & 58.332 & 52.342  \\
 \hline
\end{tabular}
}
\label{tab:categories}
\end{table}

\begin{table}[]
\footnotesize
\centering
\caption{In the validation datasets of RefCOCOg, RefCOCO+, and RefCOCO, we studied the effect of different template lengths in BSAP on mIoU for the RIS.}
\renewcommand\arraystretch{1}
\setlength\tabcolsep{8pt}
\resizebox{0.45\textwidth}{!}{
\begin{tabular}{c|ccc}
\hline 
Length & RefCOCOg & RefCOCO+ & RefCOCO\\
\hline
 0 & 33.78 & 28.12 & 27.23 \\
 1 & 33.89 & 28.40 & 27.21 \\
 2 & 33.93 & \textbf{28.86} & \textbf{27.59} \\
 3 & 33.99 & 28.67 & 27.41 \\
 4 & 34.22 & 28.56 & 26.95 \\
 5 & \textbf{34.31} & 28.53 & 26.93 \\
 6 & 34.31 & 28.48 & 26.93 \\
 7 & 34.22 & 28.48 & 26.88 \\
\hline
\end{tabular}
}
\label{tab:length}
\end{table}

\begin{table}[]
\footnotesize
\centering
\caption{In the validation datasets of RefCOCOg, RefCOCO+, and RefCOCO, we studied the effect of different balanced methods for BSAP on our best performance accuracy in the REC.}
\renewcommand\arraystretch{1.2}
\setlength\tabcolsep{2pt}
\resizebox{0.45\textwidth}{!}{
\begin{tabular}{c|ccc}
\hline 
Method & RefCOCOg & RefCOCO+ & RefCOCO\\
\hline
 Direct Normalization & 68.89 & 57.98 & 58.45 \\
 Min-max Normalization & 69.02 & 58.83 & 58.72 \\
 SoftMax & \textbf{69.65} & \textbf{59.80} & \textbf{60.90} \\
\hline   
\end{tabular}   
}
\label{tab:softmax}
\end{table}

\begin{table}[]
\footnotesize
\centering
\caption{Apply BSAP to CLIP in the image-to-text retrieval task using the CIFAR-10 dataset.}
\renewcommand\arraystretch{1}
\setlength\tabcolsep{15pt}
\resizebox{0.4\textwidth}{!}
{
\begin{tabular}{c|ccc}
\hline 
Class & CLIP & CLIP + BSAP\\
\hline
 airplane & 75 & 91 $\uparrow$ \\
 automobile & 74 & 76 $\uparrow$ \\
 bird & 78 & 81 $\uparrow$ \\
 cat & 58 & 59 $\uparrow$ \\
 deer & 54 & 94 $\uparrow$ \\
 dog & 51 & 79 $\uparrow$ \\
 frog & 63 & 92 $\uparrow$ \\
 horse & 92 & 87 \\
 ship & 92 & 96 $\uparrow$ \\
 truck & 83 & 92 $\uparrow$ \\
\hline
\end{tabular}
}
\label{tab:image-to-text}
\end{table}

\subsubsection{Effects of Contents Selection} Due to the nature of zero-shot tasks, there are no referring descriptions in datasets in advance, making it impossible to design our auxiliary prompts based on these references. For core words of auxiliary prompts, 80 categories from the COCO dataset~\cite{COCO}, 100 categories from the CIFAR dataset~\cite{CIFAR}, and 101 categories from the Caltech dataset~\cite{caltech} are tested. These datasets are widely used, and their object categories contain commonly seen objects. The impact of different quantities of auxiliary prompts is studied on the results using these categories, as shown in Table~\ref{tab:categories}. Finally, 180 categories from COCO dataset and CIFAR dataset are chosen as core words for our auxiliary prompts. The number of auxiliary prompts is also 180.

\subsubsection{Effects of Template Length} Whether the textual information input into the CLIP model is incorporated into the template can indeed impact final results. Since we are dealing with a zero-shot problem, our template selection is limited to considering the length of the text. Prompts for ChatGPT~\cite{GPT3} are used to generate templates of different lengths, e.g., ``a photo of" and ``this is". By generating ten templates for each length, fluctuation in the results is observed, as shown in Table~\ref{tab:length}. Finally, template lengths of 2, 2, and 5 are selected for RefCOCO, RefCOCO+, and RefCOCOg, respectively. 

\subsubsection{The impact of Normalization Strategies.} For balanced similarity scores generated for query text and auxiliary prompts with many image regions, we have explored several straightforward and common methods for normalization. These include direct normalization, Min-max normalization, and softmax operation. The direct normalization technique scales, mapping the values to the range of 0 to 1. Min-max normalization, also known as feature scaling, transforms data linearly to a specified range, typically [0, 1], making it suitable for various analyses and machine learning algorithms. In Table~\ref{tab:softmax}, we compared these three operations and ultimately observed that the softmax operation is superior.

\subsubsection{Analysis on Constant Parameter $\alpha$}
\begin{figure}[]
    \centering
    \subfigure[The REC task]{\includegraphics[width=0.49\linewidth]{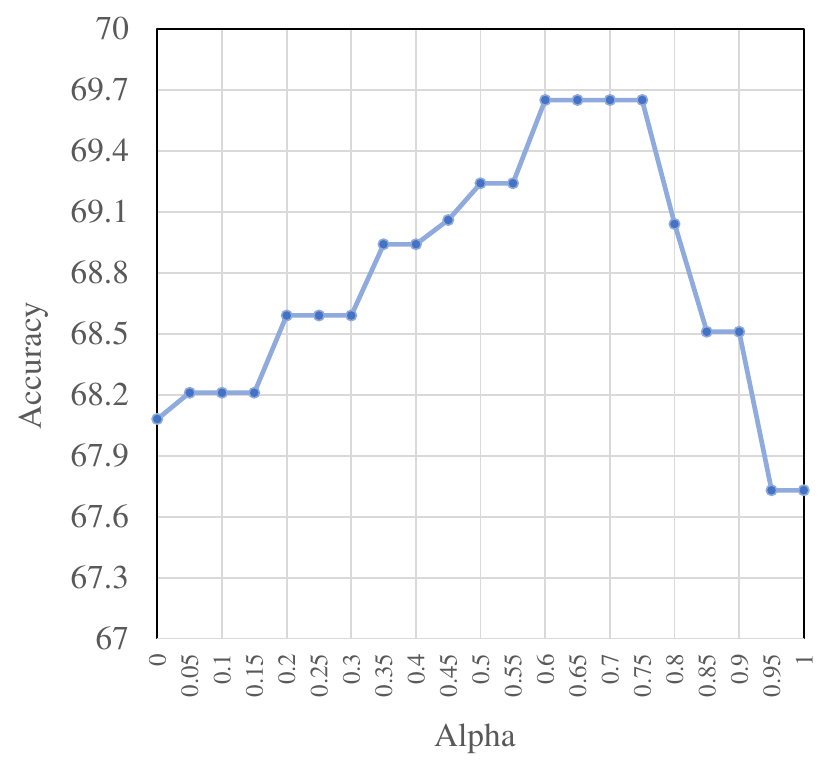}}
    \subfigure[The RIS task]{\includegraphics[width=0.49\linewidth]{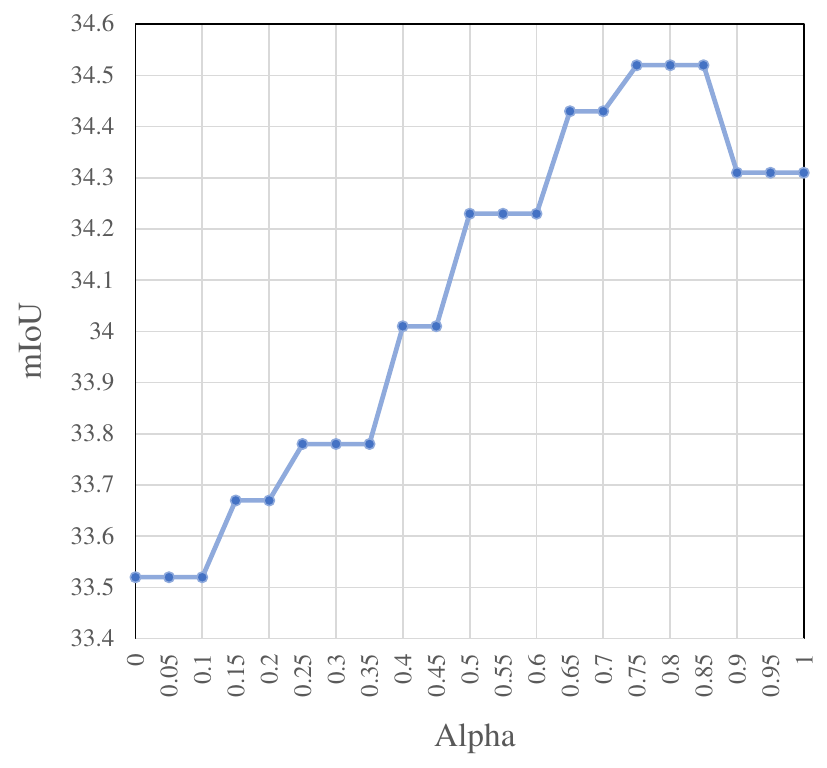}}
\caption{The impact of different $\alpha$ values on the refococog val set.(a) Different $\alpha$ values in BSAP-H on Reclip. (b) Different $\alpha$ values in BSAP-H on Zero-shot-RIS.
}
\label{fig:alpha}
\end{figure}

Since our task is zero-shot, we strive to select constant parameters $\alpha$ using as small a dataset as possible. In formula~\ref{alpha} of the hybrid method, we use $\alpha = 0.75$ for the REC task and $\alpha = 0.5$ for the RIS task in our final experiments. We conducted parameter experiments on the validation set of the RefCOCOg dataset, as shown in Fig.~\ref{fig:alpha} (a) and Fig.~\ref{fig:alpha} (b). Notably, for the RIS task, we exclusively employ ViT-B/32 as the backbone for CLIP and evaluation metric mIoU to investigate the parameter $\alpha$. For the constant parameter $\alpha$, optimal values may exist for different tasks, datasets, CLIP backbones, and evaluation metrics.

\subsection{Applying BSAP to Different VLMs (Q4)}

\begin{figure*}[ht]
\centering
\includegraphics[width=\linewidth]{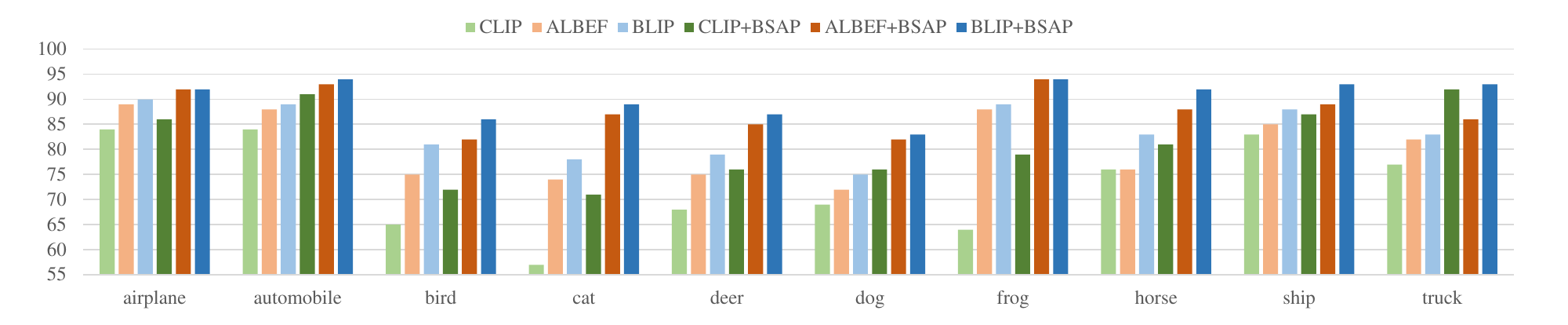}
\caption{Apply BSAP to different VLMs in the text-to-image retrieval task using the CIFAR-10 dataset. The light-colored bars represent the number of correct predictions made directly by using VLM, while the dark-colored bars represent the number of correct predictions made by integrating BSAP into VLM for prediction.}
\label{fig:bar}
\end{figure*}

We randomly sampled one image from each of the 10 classes of the CIFAR-10 dataset, using the category as our query text to see which image yielded the highest confidence score after being processed by the VLM. If the image corresponding to the highest confidence score was indeed from that category, the VLM's judgment was considered correct. Conversely, if it was not from that category, the judgment was deemed incorrect. The experiment performed the aforementioned operation 100 times for each of the 10 categories, utilizing the CLIP, ALBEF, and BLIP models. As shown in Fig.~\ref{fig:bar}, the light-colored bars indicate the number of successful judgments achieved. To demonstrate the reduction of hallucinations by our BSAP, we integrated BSAP into the three models and compared the balanced scores as shown by the dark-colored bars in Fig.~\ref{fig:bar}. Here, BSAP refers to the application of other categories within CIFAR.

As can be observed from the bar chart, the performance of the models has been enhanced and hallucinations have been mitigated after the integration of BSAP. Improvements are noted across different VLMs. 

We have also applied it to our two target tasks in Table~\ref{tab:REC_Result} and Table~\ref{tab:RIS_Result}, REC and RIS. Since both ALBEF and BLIP were trained on the COCO dataset, they cannot be considered to be making zero-shot judgments. Nevertheless, the task still demonstrates the performance improvement of ALBEF and BLIP, indicating that our BSAP has a certain mitigating effect on the hallucination phenomenon in this class of large models.

\subsection{Further Discussion of Applying BSAP in Image-to-Text Retrieval Task}
Given that our primary focus is on hallucinations in zero-shot tasks for large models, specifically within the context of the text-to-image retrieval task, we have identified that BSAP can alleviate such hallucinations. Extending our investigation to image-to-text retrieval tasks, we conducted preliminary experiments. We randomly selected one image from each category in the CIFAR-10 dataset as our query image and tasked it with identifying the corresponding CIFAR-10 category. A correct identification of the category was considered a correct judgment; otherwise, it was deemed incorrect. Using both the original CLIP and the CLIP integrated with BSAP, we performed 100 draws for each of the ten categories, as detailed in Table~\ref{tab:image-to-text}. For BSAP, an auxiliary prompt image was used, which was randomly selected from another category to serve as the BSAP. The experimental results indicate that, for the majority of categories, the use of BSAP provided significant guidance, suggesting that our BSAP also has a mitigating effect on hallucinations within image-to-text retrieval tasks.

\section{Conclusion}
In this paper, we first analyzed the text-to-image retrieval hallucination of CLIP and figured out that the imbalanced similarity score ranges cause some of the hallucinations. Then, we proposed Balanced Scores with Auxiliary Prompts (BSAP), an effective plug-in module for CLIP to the text-to-image retrieval hallucination, and boosted performance in zero-shot learning. Extensive experiments concerning REC and RIS with three datasets demonstrated the capability of our BRSP. There is still room for improvement in our BSAP, such as generating appropriate prompts depending on each text query description. Besides, other cross-modal foundation models, including ALBEF~\cite{albef} and BLIP~\cite{blip}, may also contain the image-to-text retrieval hallucination. How to apply our BRSP to other foundation models is valuable for future work. Furthermore, there also exists the image-to-text retrieval hallucination in VLMs, and analyzing the image-to-text retrieval hallucination is also one of our future works.




\bibliographystyle{IEEEtran}
\bibliography{./bibliography}

\begin{thebibliography}{10}
\providecommand{\url}[1]{#1}
\csname url@samestyle\endcsname
\providecommand{\newblock}{\relax}
\providecommand{\bibinfo}[2]{#2}
\providecommand{\BIBentrySTDinterwordspacing}{\spaceskip=0pt\relax}
\providecommand{\BIBentryALTinterwordstretchfactor}{4}
\providecommand{\BIBentryALTinterwordspacing}{\spaceskip=\fontdimen2\font plus
\BIBentryALTinterwordstretchfactor\fontdimen3\font minus \fontdimen4\font\relax}
\providecommand{\BIBforeignlanguage}[2]{{%
\expandafter\ifx\csname l@#1\endcsname\relax
\typeout{** WARNING: IEEEtran.bst: No hyphenation pattern has been}%
\typeout{** loaded for the language `#1'. Using the pattern for}%
\typeout{** the default language instead.}%
\else
\language=\csname l@#1\endcsname
\fi
#2}}
\providecommand{\BIBdecl}{\relax}
\BIBdecl

\bibitem{zero-shot-learning01}
F.~Pourpanah, M.~Abdar, Y.~Luo, X.~Zhou, R.~Wang, C.~P. Lim, X.-Z. Wang, and Q.~J. Wu, ``A review of generalized zero-shot learning methods,'' \emph{IEEE transactions on pattern analysis and machine intelligence}, 2022.

\bibitem{zero-shot-learning02}
W.~Cao, Y.~Wu, Y.~Sun, H.~Zhang, J.~Ren, D.~Gu, and X.~Wang, ``A review on multimodal zero-shot learning,'' \emph{Wiley Interdisciplinary Reviews: Data Mining and Knowledge Discovery}, vol.~13, no.~2, p. e1488, 2023.

\bibitem{tmm01}
S.~Jiang, S.~Fu, N.~Lin, and Y.~Fu, ``Pretrained models and evaluation data for the khmer language,'' \emph{Tsinghua Science and Technology}, vol.~27, no.~4, pp. 709--718, 2022.

\bibitem{tmm02}
F.~Li, Z.~Chen, and Y.~Wang, ``Hlc-keplm: Hierarchical label classification based on knowledge-enhanced pretrained language model for chinese telecom,'' in \emph{2023 4th International Conference on Intelligent Computing and Human-Computer Interaction (ICHCI)}, 2023, pp. 262--266.

\bibitem{clip}
A.~Radford, J.~W. Kim, C.~Hallacy, A.~Ramesh, G.~Goh, S.~Agarwal, G.~Sastry, A.~Askell, P.~Mishkin, J.~Clark \emph{et~al.}, ``Learning transferable visual models from natural language supervision,'' in \emph{International conference on machine learning}.\hskip 1em plus 0.5em minus 0.4em\relax PMLR, 2021, pp. 8748--8763.

\bibitem{clip_detection}
S.~Esmaeilpour, B.~Liu, E.~Robertson, and L.~Shu, ``Zero-shot out-of-distribution detection based on the pre-trained model clip,'' in \emph{Proceedings of the AAAI conference on artificial intelligence}, vol.~36, no.~6, 2022, pp. 6568--6576.

\bibitem{clip_semantic_segmentation}
Z.~Zhou, Y.~Lei, B.~Zhang, L.~Liu, and Y.~Liu, ``Zegclip: Towards adapting clip for zero-shot semantic segmentation,'' in \emph{Proceedings of the IEEE/CVF Conference on Computer Vision and Pattern Recognition}, 2023, pp. 11\,175--11\,185.

\bibitem{clip_caption}
Y.~Tewel, Y.~Shalev, I.~Schwartz, and L.~Wolf, ``Zerocap: Zero-shot image-to-text generation for visual-semantic arithmetic,'' in \emph{Proceedings of the IEEE/CVF Conference on Computer Vision and Pattern Recognition}, 2022, pp. 17\,918--17\,928.

\bibitem{image_text_retrieval}
A.~Baldrati, L.~Agnolucci, M.~Bertini, and A.~Del~Bimbo, ``Zero-shot composed image retrieval with textual inversion,'' \emph{arXiv preprint arXiv:2303.15247}, 2023.

\bibitem{reclip}
S.~Subramanian, W.~Merrill, T.~Darrell, M.~Gardner, S.~Singh, and A.~Rohrbach, ``Reclip: A strong zero-shot baseline for referring expression comprehension,'' in \emph{Proceedings of the 60th Annual Meeting of the Association for Computational Linguistics (Volume 1: Long Papers)}, 2022, pp. 5198--5215.

\bibitem{refcocog01}
S.~Kazemzadeh, V.~Ordonez, M.~Matten, and T.~Berg, ``Referitgame: Referring to objects in photographs of natural scenes,'' in \emph{Proceedings of the 2014 conference on empirical methods in natural language processing (EMNLP)}, 2014, pp. 787--798.

\bibitem{refcocog02}
J.~Mao, J.~Huang, A.~Toshev, O.~Camburu, A.~L. Yuille, and K.~Murphy, ``Generation and comprehension of unambiguous object descriptions,'' in \emph{Proceedings of the IEEE conference on computer vision and pattern recognition}, 2016, pp. 11--20.

\bibitem{refcoco}
V.~K. Nagaraja, V.~I. Morariu, and L.~S. Davis, ``Modeling context between objects for referring expression understanding,'' in \emph{Computer Vision--ECCV 2016: 14th European Conference, Amsterdam, The Netherlands, October 11--14, 2016, Proceedings, Part IV 14}.\hskip 1em plus 0.5em minus 0.4em\relax Springer, 2016, pp. 792--807.

\bibitem{zero_shot_ris}
S.~Yu, P.~H. Seo, and J.~Son, ``Zero-shot referring image segmentation with global-local context features,'' in \emph{Proceedings of the IEEE/CVF Conference on Computer Vision and Pattern Recognition}, 2023, pp. 19\,456--19\,465.

\bibitem{zero-shot-learning03}
W.~Wang, V.~W. Zheng, H.~Yu, and C.~Miao, ``A survey of zero-shot learning: Settings, methods, and applications,'' \emph{ACM Transactions on Intelligent Systems and Technology (TIST)}, vol.~10, no.~2, pp. 1--37, 2019.

\bibitem{zero-shot-learning04}
J.~Chen, Y.~Geng, Z.~Chen, I.~Horrocks, J.~Z. Pan, and H.~Chen, ``Knowledge-aware zero-shot learning: Survey and perspective,'' \emph{arXiv preprint arXiv:2103.00070}, 2021.

\bibitem{ALIGN}
C.~Jia, Y.~Yang, Y.~Xia, Y.-T. Chen, Z.~Parekh, H.~Pham, Q.~Le, Y.-H. Sung, Z.~Li, and T.~Duerig, ``Scaling up visual and vision-language representation learning with noisy text supervision,'' in \emph{International conference on machine learning}.\hskip 1em plus 0.5em minus 0.4em\relax PMLR, 2021, pp. 4904--4916.

\bibitem{video_action_localization}
J.~Nam, D.~Ahn, D.~Kang, S.~J. Ha, and J.~Choi, ``Zero-shot natural language video localization,'' in \emph{Proceedings of the IEEE/CVF International Conference on Computer Vision}, 2021, pp. 1470--1479.

\bibitem{foundation_models}
A.~Narayan, I.~Chami, L.~Orr, S.~Arora, and C.~R{\'e}, ``Can foundation models wrangle your data?'' \emph{arXiv preprint arXiv:2205.09911}, 2022.

\bibitem{resnet}
K.~He, X.~Zhang, S.~Ren, and J.~Sun, ``Deep residual learning for image recognition,'' in \emph{Proceedings of the IEEE conference on computer vision and pattern recognition}, 2016, pp. 770--778.

\bibitem{visual_transformer}
A.~Dosovitskiy, L.~Beyer, A.~Kolesnikov, D.~Weissenborn, X.~Zhai, T.~Unterthiner, M.~Dehghani, M.~Minderer, G.~Heigold, S.~Gelly \emph{et~al.}, ``An image is worth 16x16 words: Transformers for image recognition at scale,'' \emph{arXiv preprint arXiv:2010.11929}, 2020.

\bibitem{clip_limitation01}
J.-J. Shao, J.-X. Shi, X.-W. Yang, L.-Z. Guo, and Y.-F. Li, ``Investigating the limitation of clip models: The worst-performing categories,'' \emph{arXiv preprint arXiv:2310.03324}, 2023.

\bibitem{Hallucination01}
S.~Tonmoy, S.~Zaman, V.~Jain, A.~Rani, V.~Rawte, A.~Chadha, and A.~Das, ``A comprehensive survey of hallucination mitigation techniques in large language models,'' \emph{arXiv preprint arXiv:2401.01313}, 2024.

\bibitem{Hallucination02}
V.~Rawte, A.~Sheth, and A.~Das, ``A survey of hallucination in large foundation models,'' \emph{arXiv preprint arXiv:2309.05922}, 2023.

\bibitem{Hallucination03}
Q.~Huang, X.~Dong, P.~Zhang, B.~Wang, C.~He, J.~Wang, D.~Lin, W.~Zhang, and N.~Yu, ``Opera: Alleviating hallucination in multi-modal large language models via over-trust penalty and retrospection-allocation,'' \emph{arXiv preprint arXiv:2311.17911}, 2023.

\bibitem{rec}
L.~Yu, Z.~Lin, X.~Shen, J.~Yang, X.~Lu, M.~Bansal, and T.~L. Berg, ``Mattnet: Modular attention network for referring expression comprehension,'' in \emph{Proceedings of the IEEE conference on computer vision and pattern recognition}, 2018, pp. 1307--1315.

\bibitem{ris}
C.~Liu, Z.~Lin, X.~Shen, J.~Yang, X.~Lu, and A.~Yuille, ``Recurrent multimodal interaction for referring image segmentation,'' in \emph{Proceedings of the IEEE international conference on computer vision}, 2017, pp. 1271--1280.

\bibitem{visual_grounding}
C.~Zhu, Y.~Zhou, Y.~Shen, G.~Luo, X.~Pan, M.~Lin, C.~Chen, L.~Cao, X.~Sun, and R.~Ji, ``Seqtr: A simple yet universal network for visual grounding,'' in \emph{European Conference on Computer Vision}.\hskip 1em plus 0.5em minus 0.4em\relax Springer, 2022, pp. 598--615.

\bibitem{CPT}
Y.~Yao, A.~Zhang, Z.~Zhang, Z.~Liu, T.-S. Chua, and M.~Sun, ``Cpt: Colorful prompt tuning for pre-trained vision-language models,'' \emph{arXiv preprint arXiv:2109.11797}, 2021.

\bibitem{plm}
H.~Wang, J.~Li, H.~Wu, E.~Hovy, and Y.~Sun, ``Pre-trained language models and their applications,'' \emph{Engineering}, 2022.

\bibitem{GradCAM}
R.~R. Selvaraju, M.~Cogswell, A.~Das, R.~Vedantam, D.~Parikh, and D.~Batra, ``Grad-cam: Visual explanations from deep networks via gradient-based localization,'' in \emph{Proceedings of the IEEE international conference on computer vision}, 2017, pp. 618--626.

\bibitem{region_token}
J.~Li, G.~Shakhnarovich, and R.~A. Yeh, ``Adapting clip for phrase localization without further training,'' \emph{arXiv preprint arXiv:2204.03647}, 2022.

\bibitem{cropping02}
X.~Gu, T.-Y. Lin, W.~Kuo, and Y.~Cui, ``Open-vocabulary object detection via vision and language knowledge distillation,'' \emph{arXiv preprint arXiv:2104.13921}, 2021.

\bibitem{norm01}
S.~Santurkar, D.~Tsipras, A.~Ilyas, and A.~Madry, ``How does batch normalization help optimization?'' \emph{Advances in neural information processing systems}, vol.~31, 2018.

\bibitem{norm02}
J.~L. Ba, J.~R. Kiros, and G.~E. Hinton, ``Layer normalization,'' \emph{arXiv preprint arXiv:1607.06450}, 2016.

\bibitem{COCO}
T.-Y. Lin, M.~Maire, S.~Belongie, J.~Hays, P.~Perona, D.~Ramanan, P.~Doll{\'a}r, and C.~L. Zitnick, ``Microsoft coco: Common objects in context,'' in \emph{Computer Vision--ECCV 2014: 13th European Conference, Zurich, Switzerland, September 6-12, 2014, Proceedings, Part V 13}.\hskip 1em plus 0.5em minus 0.4em\relax Springer, 2014, pp. 740--755.

\bibitem{CIFAR}
A.~Krizhevsky, G.~Hinton \emph{et~al.}, ``Learning multiple layers of features from tiny images,'' 2009.

\bibitem{CoOp}
K.~Zhou, J.~Yang, C.~C. Loy, and Z.~Liu, ``Learning to prompt for vision-language models,'' \emph{International Journal of Computer Vision}, vol. 130, no.~9, pp. 2337--2348, 2022.

\bibitem{CoCoOp}
{Zhou, Kaiyang and Yang, Jingkang and Loy, Chen Change and Liu, Ziwei}, ``Conditional prompt learning for vision-language models,'' in \emph{Proceedings of the IEEE/CVF Conference on Computer Vision and Pattern Recognition}, 2022, pp. 16\,816--16\,825.

\bibitem{GPT3}
L.~Floridi and M.~Chiriatti, ``Gpt-3: Its nature, scope, limits, and consequences,'' \emph{Minds and Machines}, vol.~30, pp. 681--694, 2020.

\bibitem{hybird}
S.~Jia, C.~Ma, T.~Yao, B.~Yin, S.~Ding, and X.~Yang, ``Exploring frequency adversarial attacks for face forgery detection,'' in \emph{Proceedings of the IEEE/CVF Conference on Computer Vision and Pattern Recognition}, 2022, pp. 4103--4112.

\bibitem{miou01}
H.~Ding, C.~Liu, S.~Wang, and X.~Jiang, ``Vision-language transformer and query generation for referring segmentation,'' in \emph{Proceedings of the IEEE/CVF International Conference on Computer Vision}, 2021, pp. 16\,321--16\,330.

\bibitem{miou02}
N.~Kim, D.~Kim, C.~Lan, W.~Zeng, and S.~Kwak, ``Restr: Convolution-free referring image segmentation using transformers,'' in \emph{Proceedings of the IEEE/CVF Conference on Computer Vision and Pattern Recognition}, 2022, pp. 18\,145--18\,154.

\bibitem{uniter}
Y.-C. Chen, L.~Li, L.~Yu, A.~El~Kholy, F.~Ahmed, Z.~Gan, Y.~Cheng, and J.~Liu, ``Uniter: Universal image-text representation learning,'' in \emph{European conference on computer vision}.\hskip 1em plus 0.5em minus 0.4em\relax Springer, 2020, pp. 104--120.

\bibitem{albef}
J.~Li, R.~Selvaraju, A.~Gotmare, S.~Joty, C.~Xiong, and S.~C.~H. Hoi, ``Align before fuse: Vision and language representation learning with momentum distillation,'' \emph{Advances in neural information processing systems}, vol.~34, pp. 9694--9705, 2021.

\bibitem{blip}
J.~Li, D.~Li, C.~Xiong, and S.~Hoi, ``Blip: Bootstrapping language-image pre-training for unified vision-language understanding and generation,'' in \emph{International Conference on Machine Learning}.\hskip 1em plus 0.5em minus 0.4em\relax PMLR, 2022, pp. 12\,888--12\,900.

\bibitem{lavt}
Z.~Yang, J.~Wang, Y.~Tang, K.~Chen, H.~Zhao, and P.~H. Torr, ``Lavt: Language-aware vision transformer for referring image segmentation,'' in \emph{Proceedings of the IEEE/CVF Conference on Computer Vision and Pattern Recognition}, 2022, pp. 18\,155--18\,165.

\bibitem{freesolo}
X.~Wang, Z.~Yu, S.~De~Mello, J.~Kautz, A.~Anandkumar, C.~Shen, and J.~M. Alvarez, ``Freesolo: Learning to segment objects without annotations,'' in \emph{Proceedings of the IEEE/CVF Conference on Computer Vision and Pattern Recognition}, 2022, pp. 14\,176--14\,186.

\bibitem{tseg}
R.~Strudel, I.~Laptev, and C.~Schmid, ``Weakly-supervised segmentation of referring expressions,'' \emph{arXiv preprint arXiv:2205.04725}, 2022.

\bibitem{caltech}
M.~A. Shah, R.~Olivier, and B.~Raj, ``Exploiting non-linear redundancy for neural model compression,'' in \emph{2020 25th International Conference on Pattern Recognition (ICPR)}.\hskip 1em plus 0.5em minus 0.4em\relax IEEE, 2021, pp. 9928--9935.

\end{thebibliography}

\clearpage
\onecolumn

\end{document}